\documentclass[prc,superscriptaddress,noshowpacs,unsortedaddress,twocolumn,showpacs,preprintnumbers,amsmath,amssymb]{revtex4-1}

\usepackage[dvipdfmx]{graphicx}
\usepackage{amsmath,amssymb,times}
\usepackage{color}
\usepackage{ulem}
\usepackage{bm}
\usepackage{here}

\def\gsim{~\,\makebox(1,1){$\stackrel{>}{\widetilde{}}$}\,~}

\newcommand{\beq}{\begin{equation}}
\newcommand{\eeq}{\end{equation}}
\newcommand{\bea}{\begin{eqnarray}}
\newcommand{\eea}{\end{eqnarray}}

\newcommand{\bfi}[1]{\mbox{\boldmath $#1$}}

\newcommand{\vK}{{\bfi K}}

\newcommand{\vs}{{\bfi s}}

\newcommand{\vrr}{{\bfi r}}
\newcommand{\vR}{{\bfi R}}

\def\a{\alpha}

\begin{document}


\title{Reaction cross section of proton scattering consistent with PREX-II 
}



\author{Tomotsugu~Wakasa}

\author{Shingo~Tagami}
   
\author{Jun~Matsui}
\author{Masanobu Yahiro}
\email[]{orion093g@gmail.com}
\affiliation{Department of Physics, Kyushu University, Fukuoka 819-0395, Japan} 
\author{Maya~Takechi}
\affiliation{Niigata University, Niigata 950-2181, Japan}



\date{\today}

\begin{abstract}
\begin{description}
\item[Background]
The neutron skin thickness $R_{\rm skin}^{\rm PV}$ of PREX-II is presented in Phys. Rev. Lett. {\bf 126}, 172502 (2021). 
The reaction cross section $\sigma_R$ is useful to determine the matter radius $R_m$ and $R_{\rm skin}$.
For proton scattering, the reaction cross section $\sigma_R$ are available for $E_{\rm lab} \gsim 400$ MeV. 

\item[Method and results]
We determine $R_n^{\rm exp}=5.727 \pm 0.071$ fm and $R_m^{\rm exp}=5.617 \pm 0.044$ fm from 
$R_p^{\rm exp}$ = 5.444 fm and $R_{\rm skin}^{\rm PV}$. 
The $R_p^{\rm GHFB}$  
calculated with Gongny-D1S HFB (GHFB) with the angular momentum  projection (AMP) 
agrees with $R_p^{\rm exp}$. 
The neutron density calculated with GHFB+AMP is scaled so as to  $R_n^{\rm scaling}=5.727$ fm. 
The  Love-Franey $t$-matrix model with the scaled densities reproduces the data on $\sigma_R$. 

\item[Aim]
Our aim is to find the $\sigma_R$ of proton scattering consistent with $R_{\rm skin}^{\rm PV}$.

\item[Conclusion]
The $\sigma_R$ of proton scattering consistent with $R_{\rm skin}^{\rm PV}$ are 
$\sigma_R^{\rm exp}$ at $E_{\rm lab} = 534.1, 549, 806$ MeV.

\end{description}
\end{abstract}


\maketitle

\clearpage

\section{Introduction and conclusion}
\label{Introduction}

{\bf Background:}
Horowitz {\it et al.} \cite{PRC.63.025501} proposed a direct measurement 
for neutron skin $R_{\rm skin}$. 
The measurement is composed of parity-violating ($PV$) weak scattering and 
elastic electron scattering. 
The neutron radius $R_n$ is determined from the former experiment, whereas
the proton radius $R_p$ is from the latter.

Very recently, the PREX collaboration presented the PREX-II value~\cite{Adhikari:2021phr}:
\begin{equation}
R_{\rm skin}^{PV} = 0.283\pm 0.071\,{\rm fm},
\end{equation}
combining the original Lead Radius EXperiment (PREX) 
result \cite{PRL.108.112502,PRC.85.032501} with the updated PREX-II result.
 The $R_{\rm skin}^{PV}$ value is most reliable at the present 
stage, and provides crucial tests for the equation of state (EoS) 
of nuclear matter 
\cite{PRC.102.051303,AJ.891.148,AP.411.167992,EPJA.56.63,JPG.46.093003}
as well as nuclear structure and reaction.  In particular, Reed {\it et al.} \cite{arXiv.2101.03193} 
report a value of the slope parameter of the EoS 
and examine the impact of such a stiff symmetry energy 
on some critical neutron-star observables.
The $R_{\rm skin}^{PV}$ value 
is considerably larger than the other experimental 
values which are significantly model dependent 
\cite{PRL.87.082501,PRC.82.044611,PRL.107.062502,%
PRL.112.242502}.
The nonlocal dispersive-optical-model 
(DOM) analysis of ${}^{208}{\rm Pb}$ deduces 
$r_{\rm skin}^{\rm DOM} =0.25 \pm 0.05$ fm \cite{PRC.101.044303}, 
The chiral (Kyushu) $g$-matrix folding model  
determines $r_{\rm skin}^{208}=0.27 \pm 0.03$~fm from reaction cross section 
$\sigma_{\rm R}$ in $30 \leq E_{\rm lab} \leq 100$~MeV~\cite{Tagami:2020bee}. 
These values are consistent with $R_{\rm skin}^{PV}$.

{\bf Aim:}
The aim is to find the $\sigma_R$ of $p+{}^{208}{\rm Pb}$ scattering 
that supports $R_{\rm skin}({\rm PREXII})$.

{\bf Method and results:}
The reaction cross section $\sigma_R$ is a powerful tool of evaluating the matter  radius $R_m$. 
We first determine $R_n^{\rm exp}=5.727 \pm 0.071$ fm and $R_m^{\rm exp}=5.617 \pm 0.044$ fm 
from $R_p^{\rm exp}$ = 5.444 fm \cite{PRC.90.067304} and $R_{\rm skin}^{\rm PV}$. 
The $R_p^{\rm GHFB}$  calculated with Gongny-D1S HFB (GHFB) 
with the angular momentum  projection (AMP) agrees with $R_p^{\rm exp}$ of electron scaling. 
The neutron density calculated with GHFB+AMP is scaled so as to  $R_n^{\rm scaling}=5.727$ fm. 
The  Love-Franey $t$-matrix~\cite{LF} model with the scaled densities reproduces 
the data on $\sigma_R$ at $E_{\rm lab} = 534.1, 549, 806$ MeV. 
Our calculation has no free parameter. 

{\bf Conclusion:}
The $\sigma_R$ of proton scattering consistent with $R_{\rm skin}^{\rm PV}$ are 
$\sigma_R^{\rm exp}$ at $E_{\rm lab} = 534.1, 549, 806$ MeV.

\section{Model}
\label{sec:model}

 Our model is the folding model  based on Love-Franey (LF) $t$-matrix~\cite{LF}.

 The formulation of the folding model is shown below. 
 For proton-nucleus scattering, the potential $U(\vR)$ 
 between an incident proton  and a target (${\rm T}$) has the direct and exchange parts,
$U^{\rm DR}$ and $U^{\rm EX}$, as
\begin{subequations}
\begin{eqnarray}
U^{\rm DR}(\vR) & = & 
\sum_{\nu}\int             \rho^{\nu}_{\rm T}(\vrr_{\rm T})
            t^{\rm DR}_{\mu=-1/2,\nu}(s)  d
	    \vrr_{\rm T}\ ,\label{eq:UD} \\
U^{\rm EX}(\vR) & = & 
\sum_{\nu}
\int \rho^{\nu}_{\rm T}(\vrr_{\rm T},\vrr_{\rm T}+\vs) \nonumber \\
                &   &
\times t^{\rm EX}_{\mu=-1/2,\nu}(s) \exp{[-i\vK(\vR) \cdot \vs/M]}
             d \vrr_{\rm T}\,~~~
             \label{eq:UEX}
\end{eqnarray}
\end{subequations}
where $\vR$ is the relative coordinate between an incident proton  and T,
$\vs=-\vrr_{\rm T}+\vR$, and $\vrr_{\rm T}$ is
the coordinate of the interacting nucleon from T.
 Each of $\mu$ and $\nu$ denotes the $z$-component of isospin; 
$1/2$ means neutron and $-1/2$ does proton.
 The nonlocal $U^{\rm EX}$ has been localized in Eq.~\eqref{eq:UEX}
with the local semi-classical approximation
\cite{NPA.291.299,*NPA.291.317,*NPA.297.206},
where \vK(\vR) is the local momentum between an incident proton and T, 
and $M= A/(1 +A)$ for the target mass number $A$;
see Ref.~\cite{JPG.37.085011} for the validity of the localization.

The direct and exchange parts, $t^{\rm DR}_{\mu\nu}$ and 
$t^{\rm EX}_{\mu\nu}$, of the $t$ matrix are described as
\begin{align}
&\hspace*{0.5cm} t_{\mu\nu}^{\rm DR}(s) \nonumber \\ 
&=
\begin{cases}
\displaystyle{\frac{1}{4} \sum_S} \hat{S}^2 t_{\mu\nu}^{S1}
 (s) \hspace*{0.42cm} ; \hspace*{0.2cm} 
 {\rm for} \hspace*{0.1cm} \mu+\nu = \pm 1 
 \vspace*{0.2cm}\\
\displaystyle{\frac{1}{8} \sum_{S,T}} 
\hat{S}^2 t_{\mu\nu}^{ST}(s), 
\hspace*{0.2cm} ; \hspace*{0.2cm} 
{\rm for} \hspace*{0.1cm} \mu+\nu = 0 
\end{cases}
\\
&\hspace*{0.5cm}
t_{\mu\nu}^{\rm EX}(s) \nonumber \\
&=
\begin{cases}
\displaystyle{\frac{1}{4} \sum_S} (-1)^{S+1} 
\hat{S}^2 t_{\mu\nu}^{S1} (s) 
\hspace*{0.34cm} ; \hspace*{0.2cm} 
{\rm for} \hspace*{0.1cm} \mu+\nu = \pm 1 \vspace*{0.2cm}\\
\displaystyle{\frac{1}{8} \sum_{S,T}} (-1)^{S+T} 
\hat{S}^2 t_{\mu\nu}^{ST}(s) 
\hspace*{0.2cm} ; \hspace*{0.2cm}
{\rm for} \hspace*{0.1cm} \mu+\nu = 0 ~~~~~
\end{cases}
\end{align}
where $\hat{S} = {\sqrt {2S+1}}$ and $t_{\mu\nu}^{ST}$ are 
the spin-isospin components of the $t$-matrix interaction.
We apply the LF $t$-matrix  folding model for  p+$^{208}$Pb scaling  
in $E_{\rm lab}=460, 534.1, 549, 806 $~MeV.

 The relative wave function $\psi$ is decomposed into partial waves $\chi_L$,
each with different orbital angular momentum $L$.
 The elastic $S$-matrix elements $S_L$ are obtained 
from the asymptotic form of the $\chi_L$.
 The total reaction cross section $\sigma_{\rm R}$ is calculable 
from the $S_L$ as
\bea
\sigma_{\rm R}=\frac{\pi}{K^2}\sum_L (2L+1)(1-|S_L|^2)\ ,
\eea
where $\hbar K$ is an incident momentum.

As proton and neutron densities, $\rho^{\nu=-1/2}_{\rm T}$ and $\rho^{\nu=1/2}_{\rm T}$, 
 we use the densities calculated with GHFB+AMP \cite{PRC.101.014620}. 
 As a way of taking the center-of-mass correction to the densities, 
we use the method of Ref.~\cite{PRC.85.064613}, since the procedure is quite simple. 
The $R_p^{\rm GHFB}$  calculated with GHFB+AMP agrees with $R_p^{\rm exp}$= 5.444 fm \cite{PRC.90.067304}. 
The neutron density calculated with GHFB+AMP is scaled so as to  
$R_n^{\rm scaling}=5.727$ fm (the central value of $R_n^{\rm exp}=5.727 \pm 0.071$ fm determined in 
Sec.~\ref{Introduction}). 
The scaled densities based on $R_{\rm skin}^{PV}$  and $R_p^{\rm exp}$ are used for analyses of 
p+$^{208}$Pb scattering. 

Now we explain the scaling of density $\rho(\vrr)$.  
We can obtain the scaled density $\rho_{\rm scaling}(\vrr)$ from the original density $\rho(\vrr)$ as
\bea
\rho_{\rm scaling}(\vrr)=\frac{1}{\a^3}\rho(\vrr/\a)
\eea
with a scaling factor
\bea
\a=\sqrt{ \frac{\langle \vrr^2 \rangle_{\rm scaling}}{\langle \vrr^2 \rangle}} .\eea

\section{Results}
\label{sec:results} 

\begin{figure}[H]
\begin{center}
\includegraphics[width=0.5\textwidth,clip]{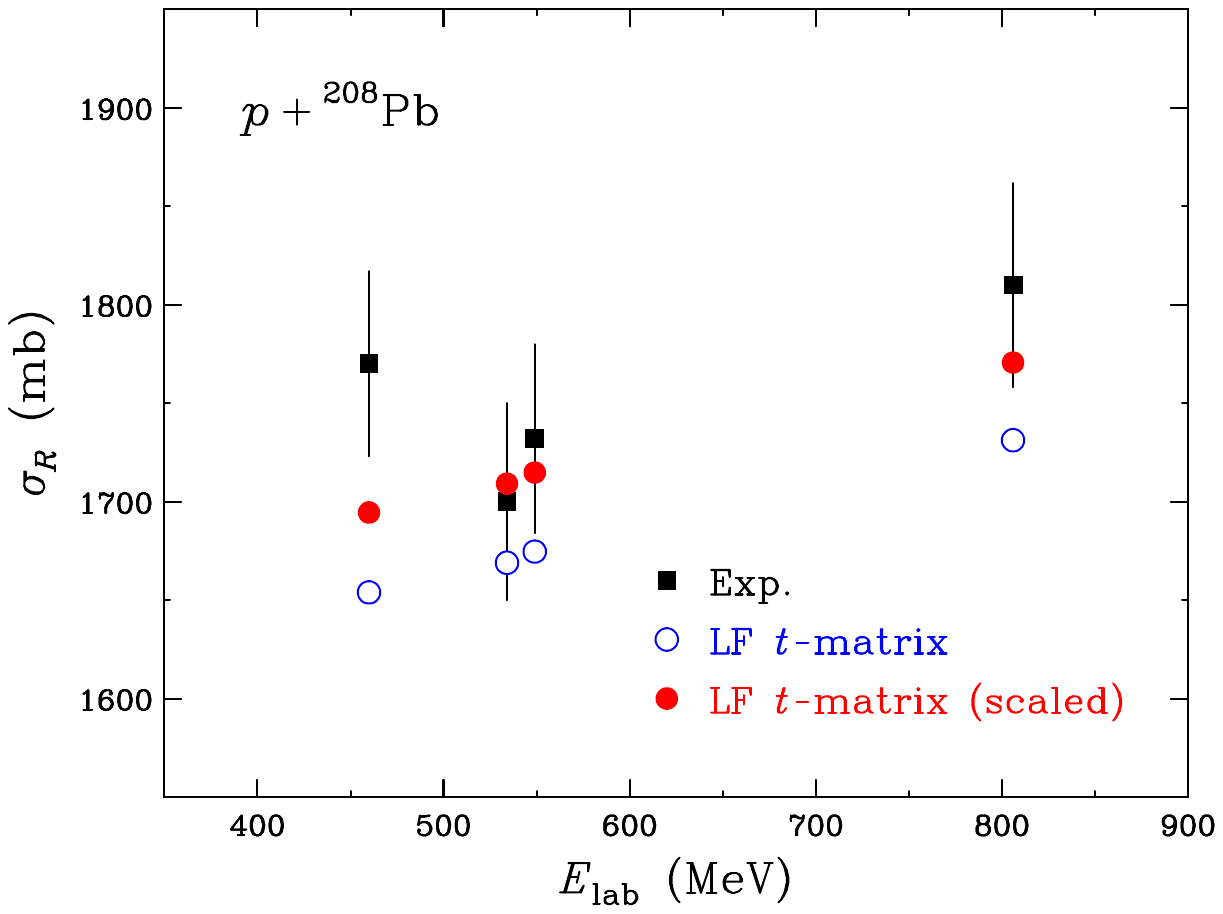}
 \caption{ 
 $E_{\rm lab}$ dependence of reaction cross sections $\sigma_{\rm R}$ 
 for $p$+$^{208}$Pb scattering. 
 Open circles stand for the results of the LF $t$-matrix  folding model with GHFB+AMP densities, 
 whereas closed circles correspond to  those of the LF $t$-matrix  folding model with the scaled 
 densities.  The data are taken from Refs.~\cite{Dietrich:2002swm,Nakano:2021dau}. 
  }
 \label{Fig-RXsec-p+Pb}
\end{center}
\end{figure}

The LF $t$-matrix folding model with the GHFB+AMP densities 
underestimates the $\sigma_R$ data 
in $400 \leq E_{\rm lab} \leq 900$~MeV only 
by a factor of 0.96, as shown in Fig.~\ref{Fig-RXsec-p+Pb}.  
The LF $t$-matrix  folding model with the scaled  densities reproduces the data in 
$E_{\rm lab}= 534.1, 549, 806$~MeV. 
This indicates that the LF $t$-matrix  folding model with the scaled  densities is useful 
in $400 \leq E_{\rm lab} \leq 900$~MeV.

\section{Discussions}
\label{sec:Discussions} 

Now we discuss how good the LF $t$-matrix  folding model with the scaled 
 densities is for p+$^{12}$C scattering at $E_{\rm lab}= 800$~MeV and  
p+$^{40}$Ca scattering at $E_{\rm lab}= 700$~MeV.

For $^{40}$Ca, Zenihiro {\it et al.} determined neutron radius $R_n({\rm RCNP})=3.375$~fm, 
$R_p({\rm RCNP})=3.385$~fm and $R_{\rm skin}({\rm RCNP})=-0.01 \pm 0.049$~fm 
from the differential cross section and the analyzing powers for p+$^{40}$Ca scattering~\cite{Zenihiro:2018rmz}.  
The GHFB+AMP densities are scaled so as to $R_p({\rm scaling})=R_p({\rm RCNP})$
and $R_n({\rm scaling})=R_n({\rm RCNP})$. 

For $^{12}$C, Tanihata {\it et al.} determined matter radius $R_{m}(\sigma_{\rm I})=2.35(2) $~fm 
from interaction cross sections $\sigma_{\rm I}$~\cite{Ozawa:2001hb}. 
We deduce neutron radius $R_{n}(\sigma_{\rm I})=2.37$~fm from 
the $R_{m}(\sigma_{\rm I})$ and the $R_{p}^{\rm exp}=2.33$~fm of electron scattering. 
The GHFB+AMP densities are scaled so as to $R_p({\rm scaling})=R_p^{\rm exp}$
and $R_n({\rm scaling})=R_n(\sigma_{\rm I})$.

Figure \ref{Fig-RXsec-p+C,Ca} shows $\sigma_{\rm R}$  
for p+$^{40}$Ca scattering at 700~MeV and  p+$^{12}$C scattering at 800~MeV. 
The LF $t$-matrix  folding model with the scaled 
 densities is good for p+$^{40}$Ca scattering at $E_{\rm lab}= 700$~MeV, 
 and almost reproduces the data for p+$^{12}$C scattering at $E_{\rm lab}= 800$~MeV.

\begin{figure}[H]
\begin{center}
\includegraphics[width=0.5\textwidth,clip]{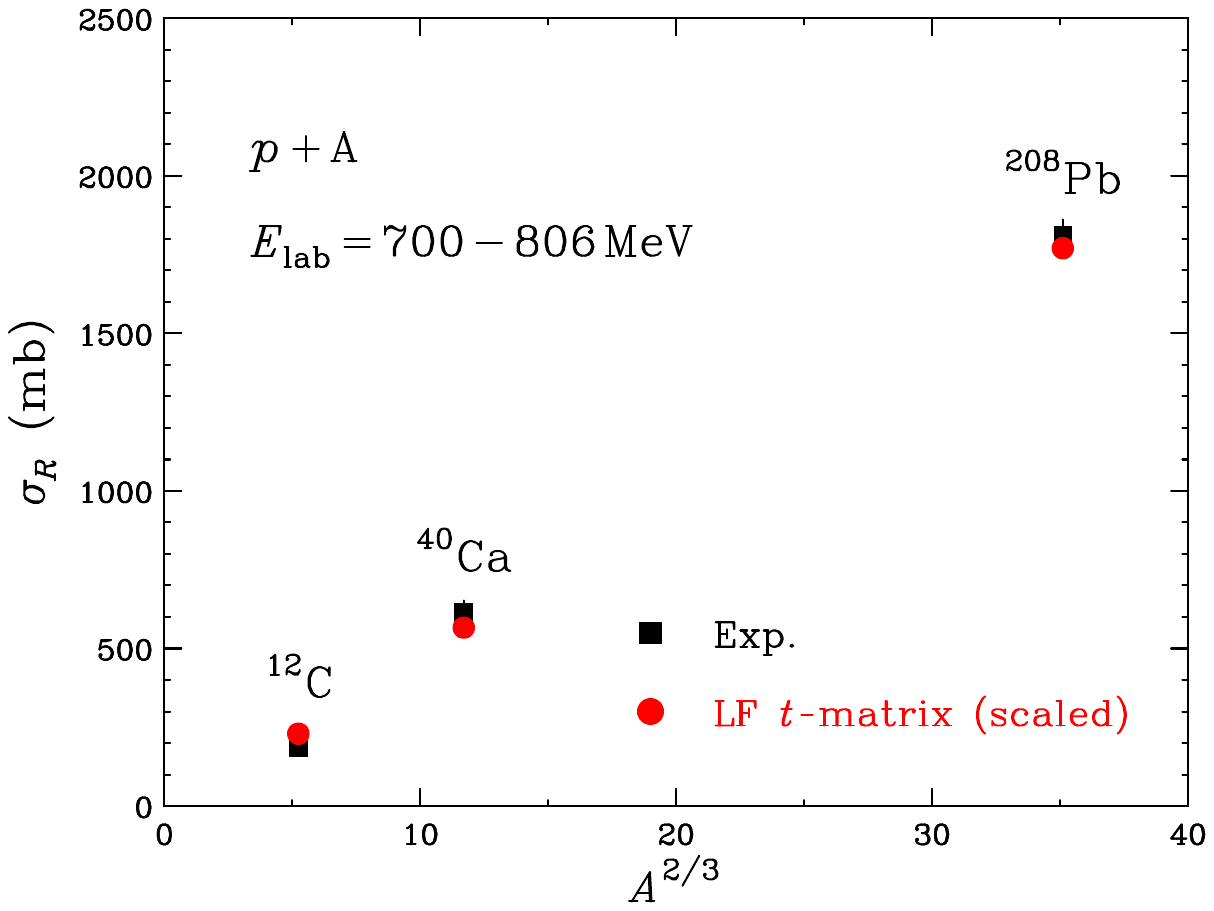}
 \caption{ 
 Mass-number $A$ dependence of reaction cross sections $\sigma_{\rm R}$ 
 for p+$^{40}$Ca scattering at 700~MeV and  p+$^{12}$C scattering at 800~MeV. 
 Closed circles stand for the results of the LF $t$-matrix  folding model with the scaled 
 densities.  The data are taken from Refs.~\cite{Anderson:1979ge,Trzaska:1991ww}. 
  }
 \label{Fig-RXsec-p+C,Ca}
\end{center}
\end{figure}

\begin{acknowledgments}
We would like to thank Dr. Toyokawa for providing his code and Prof. M. Nakano for useful information. 
\end{acknowledgments}

\bibliography{Folding-Pb}

\end{document}